\documentclass{article}

\PassOptionsToPackage{numbers, compress}{natbib}
\usepackage[preprint]{neurips_2025}

\usepackage[utf8]{inputenc} 
\usepackage[T1]{fontenc}    
\usepackage{hyperref}       
\usepackage{url}            
\usepackage{booktabs}       
\usepackage{amsfonts}       
\usepackage{nicefrac}       
\usepackage{microtype}      
\usepackage{xcolor}         
\usepackage{graphicx}
\usepackage{amssymb}
\usepackage{multirow}
\usepackage[normalem]{ulem}
\useunder{\uline}{\ul}{}
\usepackage{makecell}
\usepackage{enumitem}
\usepackage{amsmath}

\title{Reason-to-Recommend: Using Interaction-of-Thought Reasoning to Enhance LLM Recommendation}

\author{
~Keyu Zhao, Fengli Xu\footnotemark[1]\thanks{Fengli Xu is corresponding author. Email: \texttt{fenglixu@tsinghua.edu.cn}}, Yong Li   
\\
~~Department of Electronic Engineering, Tsinghua University \\
~~Beijing, China
}

\begin{document}

\maketitle

\begin{abstract}
Driven by recent advances in Large Language Models (LLMs), their integration into recommendation tasks has attracted growing interest, owing to their strong semantic understanding and prompt-based flexibility. Prior studies have encoded user–item interactions or metadata into prompts to enable recommendations. In parallel, LLM reasoning capabilities, empowered by test-time scaling technologies and reinforcement learning, have achieved remarkable success in domains such as mathematics and code, where well-defined reasoning traces and objective correctness signals are available, enabling both high performance and strong interpretability of LLM's outputs. However, directly transferring these reasoning techniques to recommendation tasks has proven ineffective, due to the implicit nature of user feedback and the lack of supervision for reasoning over interaction data. To bridge this gap, we propose a reasoning-enhanced recommendation framework, \textbf{R2Rec}, which introduces interaction chains sampled from the user–item graph and transforms them into structured interaction-of-thoughts using a progressive, masked prompting strategy, where each thought represents a stepwise reasoning grounded in user–item interaction context. This transformation enables LLMs to simulate step-by-step decision-making based on implicit interaction patterns. To internalize this reasoning ability, we design a two-stage training pipeline to improve the LLM's recommendation capability: supervised fine-tuning imparts basic reasoning skills from high-quality annotated traces, while reinforcement learning provides scalable supervision to refine the reasoning process based on reward signals, further refining the reasoning process and alleviating the scarcity of explicit interaction-reasoning data. Experimental results on three real-world datasets demonstrate the effectiveness of our approach: it outperforms both classical and LLM-based baselines with an average \textbf{10.48\%} improvement in HitRatio@1, and achieves an average \textbf{131.81\%} performance gain over the original LLM. Furthermore, the explicit reasoning chains offer enhanced interpretability by revealing the LLM’s decision process. Our code is open-source at \url{https://anonymous.4open.science/r/R2Rec-7C5D} for reproducibility.
\end{abstract}

\section{Introduction}
\label{introduction}
The integration of Large Language Models (LLMs) into recommendation tasks has recently garnered substantial research interest~\cite{Bao_2023,Dai_2023,geng2023recommendationlanguageprocessingrlp,liao2024llaralargelanguagerecommendationassistant,liu2023chatgptgoodrecommenderpreliminary,liu2025improvingllmpoweredrecommendationspersonalized,Zhao_2024,zhu2025collaborativeretrievallargelanguage,shang2025agentrecbench,yan2025agentsociety,tang2025thinkrecommendunleashinglatent,long2025got4recgraphthoughtssequential,Yue_Yin_Zhang_Shi_Liang_Wan_2025,yang2024chain}, owing to their strong semantic understanding, contextual reasoning capabilities, and extensive world knowledge acquired through large-scale pretraining. In parallel, enhancing the reasoning capabilities of LLMs has emerged as a key objective~\cite{muennighoff2025s1simpletesttimescaling,wen2025lightr1curriculumsftdpo,liu20251bllmsurpass405b,ye2025limoreasoning,xu2025towards}, with increasing efforts devoted to equipping LLMs with the ability to perform structured, multi-step inference, which is a property that not only improves task performance but also brings enhanced interpretability by making the decision-making process more transparent.

LLM-based recommendation approaches typically involve either adapting LLMs as standalone recommenders through prompt tuning~\cite{Zhang_2023,SHEN2023103139,yang2021improvingconversationalrecommendationsystems} and fine-tuning~\cite{Bao_2023,zhang2023recommendationinstructionfollowinglarge,li2023personalizedpromptlearningexplainable,lin2024bridgingitemslanguagetransition} or leveraging them to enrich traditional models with semantic features derived from textual metadata.  Concurrently, advances in LLM reasoning, which is driven by reinforcement learning~\cite{yu2025dapoopensourcellmreinforcement,yue2025vapoefficientreliablereinforcement,shao2024deepseekmathpushinglimitsmathematical,hu2025reinforceefficientrlhfalgorithm}, have enabled complex multi-step inference primarily in structured domains like mathematics and programming.

Despite significant progress in both LLM-based recommendation and LLM reasoning, effectively integrating robust reasoning capabilities into recommendation remains a largely underexplored challenge. Directly applying existing Large Reasoning Model (LRM, such as OpenAI o1~\cite{jaech2024openai}, Deepseek-R1~\cite{guo2025deepseek}) to recommendation tasks often yields suboptimal results, largely because these LLMs are primarily developed for domains like mathematics and code, which provide abundant and explicit supervision. In contrast, recommendation involves implicit user behaviors and lacks explicit reasoning data or supervision in natural language, creating a critical gap that motivates the need for new approaches capable of eliciting reasoning over user-item interactions.

To effectively equip LLMs with the ability to reason over user–item interactions and make informed recommendations, we propose a reasoning-augmented framework that transforms local interaction structures into interpretable reasoning traces and internalizes this reasoning ability through post-training. Our approach consists of three key components. First, we construct interaction chains by sampling from a user’s neighborhood in the user–item interaction graph and transform these chains into structured reasoning chains via a progressive, masked prompting strategy. This process enables the LLM to explicitly capture and represent complex interaction patterns in a stepwise reasoning format, producing a high-quality reasoning chain to explain the interaction between sampled users and items, i.e., Interaction-of-thought. Second, to teach the LLM how to autonomously generate such reasoning chains, we design a supervised fine-tuning (SFT) stage using high-quality annotated reasoning traces, which serves as a warm start by enabling the LLM to learn step-by-step inference over user-item data. Finally, recognizing the limited availability and high cost of supervised data, we introduce a reinforcement learning (RL) stage~\cite{luong2024reftreasoningreinforcedfinetuning,shao2024deepseekmathpushinglimitsmathematical} that provides scalable supervision through reward signals, allowing the LLM to further refine and self-improve its reasoning and recommendation capabilities beyond the SFT initialization.

We evaluate our method on three real-world recommendation datasets~\cite{10.1145/2827872,hou2024bridging}, achieving an average improvement of \textbf{10.48\%} in HitRatio@1 over state-of-the-art baselines. This performance gain stems from the LLM’s enhanced ability to reason over interaction chains. Our reasoning-augmented training pipeline substantially enhances LLM performance in recommendation tasks by effectively aligning their reasoning capabilities with task objectives. This is achieved through the integration of structured interaction representations and explicit reasoning supervision. Comprehensive ablation studies demonstrate the critical role of each component—interaction chains, reasoning transformation, supervised fine-tuning, and reinforcement learning—showing that each stage contributes uniquely and significantly to the overall improvement.

In summary, the main contributions of this work include:
\begin{itemize}[leftmargin=15pt]
    \setlength{\itemsep}{0pt}
    \item We propose a novel reasoning-augmented recommendation framework that leverages user-item interaction chains transformed into structured reasoning chains via a progressive, masked prompting strategy, significantly unlocking the reasoning capabilities of LLMs and enabling them to model user preferences more accurately and interpretably.
    \item We develop a two-stage post-training pipeline that first employs supervised fine-tuning to teach the LLM step-by-step reasoning over interaction chains, and then introduces reinforcement learning to enable self-improvement through reward-driven optimization, effectively overcoming the limitations of scarce annotated reasoning data.
    \item We conduct experiments on three widely used datasets from different domains of recommendation. Our method consistently outperforms both classical and LLM-based methods, achieving an average improvement of 10.48\% HitRatio@1.
\end{itemize}
\section{Related Works}
\label{related works}
\subsection{Graph-based recommendation}
Graph-based methods have demonstrated strong potential in recommendation by modeling user–item interactions as graphs, capturing both direct and high-order relationships~\cite{Wang_2019,10.1145/3240323.3240381}. A key approach leverages Graph Neural Networks (GNNs) such as GraphSAGE~\cite{hamilton2018inductiverepresentationlearninglarge}, NGCF~\cite{Wang_2019}, and LightGCN~\cite{he2020lightgcnsimplifyingpoweringgraph} to aggregate neighborhood information for enhanced user and item embeddings. These techniques have also been extended to sequential recommendations, modeling interaction sequences as session or dynamic graphs~\cite{9405450,chang2023sequentialrecommendationgraphneural,zhang2021dynamicgraphneuralnetworks,info11080388}. While effective, GNN-based methods lack inherent reasoning capabilities and interpretability. Inspired by these approaches, we instead extract interaction chains as structured reasoning paths, enabling LLMs to conduct dynamic, explainable recommendations through sequence-aware reasoning.
\subsection{LLM reasoning}
As Large Language Models (LLMs) are increasingly deployed across various domains~\cite{fan2025invisible,chen2024large,hao2024hlm}, recent advances demonstrate that they can significantly enhance reasoning capabilities through reinforcement learning (RL)\cite{yu2025dapoopensourcellmreinforcement,yue2025vapoefficientreliablereinforcement,shao2024deepseekmathpushinglimitsmathematical,hu2025reinforceefficientrlhfalgorithm,xu2025towards} and test-time scaling techniques such as Monte Carlo Tree Search (MCTS)\cite{choi-etal-2023-kcts,zhou2024languageagenttreesearch,zhang2023planninglargelanguagemodels,liu2024dontthrowawayvalue} and multi-path sampling~\cite{levi2024simplemodelinferencescaling,wang2023selfconsistencyimproveschainthought,brown2024largelanguagemonkeysscaling}. RL-based post-training methods, exemplified by DeepSeek-R1~\cite{guo2025deepseek} and Kimi k1.5~\cite{team2501kimi}, facilitate emergent multi-step reasoning behaviors including planning and self-verification, even without supervised initialization. Hybrid approaches like ReFT integrate supervised fine-tuning with RL to further refine reasoning quality. Despite these advances, research on LLM reasoning predominantly targets structured domains such as mathematics and programming, leaving recommendation systems—characterized by implicit user behavior and scarce explicit supervision—largely unexplored, highlighting a critical area for future investigation.
\subsection{LLM-based recommendation}
Recent advances in Large Language Models (LLMs) have spurred interest in their application to recommendation tasks, with approaches generally falling into two categories. One treats LLMs as standalone recommenders, adapting them through tuning~\cite{Bao_2023, geng2023recommendationlanguageprocessingrlp, cui2022m6}, prompting~\cite{Dai_2023}, or in-context learning~\cite{hou2024large,liu2023chatgptgoodrecommenderpreliminary}. Notable examples include ChatRec~\cite{gao2023chatrecinteractiveexplainablellmsaugmented}, which reformulates user profiles and interactions as prompts; TALLRec~\cite{Bao_2023}, fine-tuning LLMs with instruction-style recommendation data; and LLaRA~\cite{liao2024llaralargelanguagerecommendationassistant}, which treats interaction sequences as additional modalities via curriculum tuning. The other category employs LLMs as enhancers~\cite{hou2023learningvectorquantizeditemrepresentation,hou2022universalsequencerepresentationlearning,yuan2023recommendersystemsidvs}, leveraging them to generate or extract semantic features like item descriptions to augment traditional recommenders. While effective at integrating external knowledge and semantics, these methods often neglect structured reasoning over user-item interactions and rely on domain-specific pretraining or single-step prediction frameworks.

\section{Methods}
\label{method}
\subsection{Overview}
\begin{figure}[ht]
\begin{center}
\includegraphics[width=1.0\linewidth]{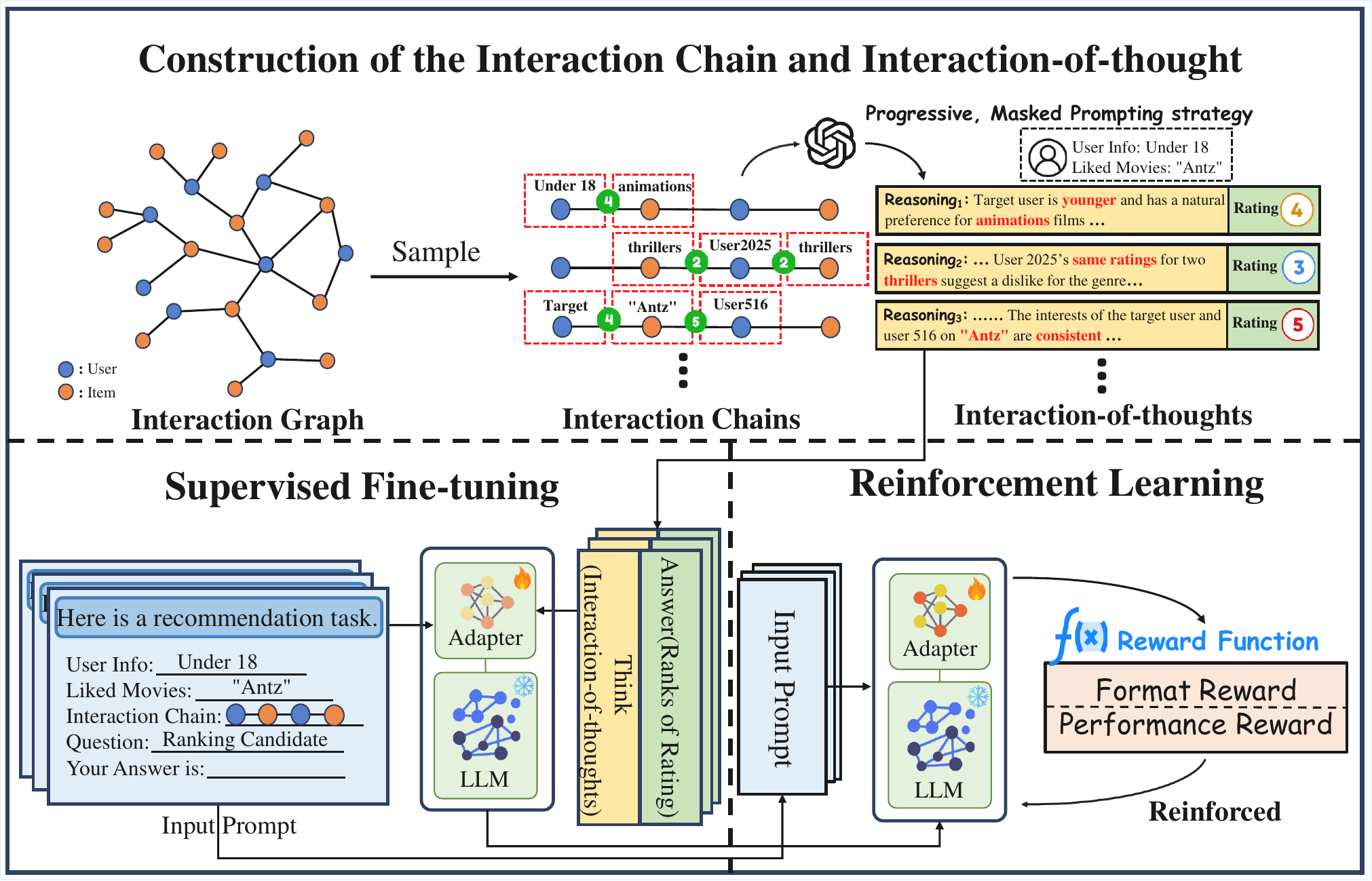}
\caption{Overview of Reason-to-Recommend(R2Rec) Framework.}   
\label{Fig1}
\end{center}
\end{figure}
To enhance the reasoning ability of LLMs in recommendation tasks, we propose a method to enhance the reasoning capabilities of LLMs for recommendation by leveraging structured user-item interaction data. We first construct interaction chains by sampling from the neighborhood of a target user within the interaction graph. These chains are then transformed into reasoning chains using a progressive, masked prompting strategy, allowing the LLM to perform step-by-step inference. To internalize this reasoning ability, we adopt a two-stage post-training approach: supervised fine-tuning provides an initial understanding of reasoning over interaction chains, while reinforcement learning enables scalable self-improvement through reward-driven optimization. We illustrate the overview of our method in Figure~\ref{Fig1}.

\subsection{Transforming Interaction Chain into Interaction-of-Thought}
\begin{figure}[ht]
\begin{center}
\includegraphics[width=1.0\linewidth]{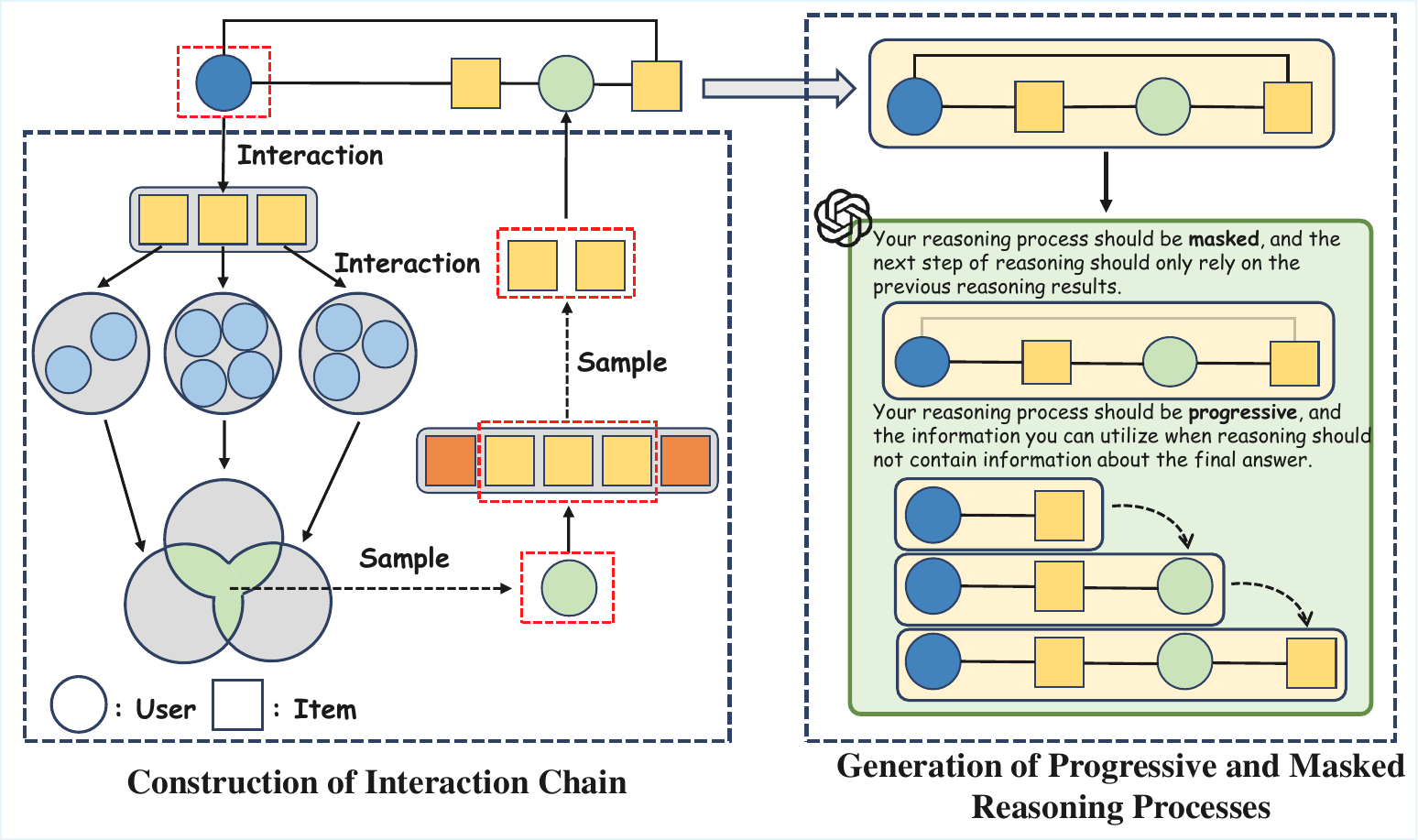}
\caption{The construction of the interaction chain and the progressive and masked prompt strategies used to generate the reasoning process.}   
\label{Fig2}
\end{center}
\end{figure}
User-item interaction graphs capture rich structural patterns that reflect users’ preferences through their local neighborhoods. However, directly encoding such graphs into LLMs is impractical due to input length limitations and the LLMs’ difficulty in processing complex topologies. To address this, we extract interaction chains by sampling from the neighborhood of a target user, forming short sequences that retain meaningful behavioral context while remaining suitable for prompt-based input. To further enhance the interpretability and reasoning potential of these chains, we transform them into step-by-step reasoning chains using a progressive, masked prompting strategy, enabling LLMs to infer preferences through structured, natural language reasoning. This process is illustrated in Figure~\ref{Fig2}.
\label{construction}
\subsubsection{Interaction Chain}
Interaction chains are constructed to capture meaningful local structures within the user–item interaction graph, serving as foundational inputs for downstream reasoning and recommendation by LLMs. However, designing effective chains is non-trivial—naive sampling may produce sequences where the final item is unrelated to the target user, leading to weak preference signals. To address these challenges, we adopt a neighborhood sampling strategy with a closure constraint that ensures the interaction chain forms a valid behavioral loop. Concretely, given a target user $u_0$, we first collect their interacted item set $\mathcal{I}{u_0} = { i_1, i_2, \dots, i_n }$. For each item $i_k$ in this set, we retrieve the users who have also interacted with it, denoted as $\mathcal{U}{i_k} = { u \mid (u, i_k) \in \mathcal{E} }$. To identify candidate users exhibiting similar preferences, we define:
\begin{equation}
    \mathcal{U}_{\text{cand}} = \bigcup_{\substack{i_p, i_q \in \mathcal{I}_{u_0} \\ i_p \ne i_q}} ( \mathcal{U}_{i_p} \cap \mathcal{U}_{i_q} ) \setminus \{ u_0 \}
\end{equation}
From this candidate set, we randomly sample an intermediate user $u_1$. Next, we determine the shared item set $\mathcal{I}_{\text{shared}} = \mathcal{I}{u_0} \cap \mathcal{I}_{u_1}$, from which two items $i_0, i_1$ are selected to form the interaction chain: $u_0 \rightarrow i_0 \rightarrow u_1 \rightarrow i_1 \rightarrow u_0$. This construction guarantees that the chain both begins and ends with interactions involving the target user $u_0$, thereby preserving behavioral coherence and grounding subsequent reasoning in authentic user-item interactions to support accurate preference modeling.
\subsubsection{Interaction-of-thoughts}
\label{reasoning chain}

To enable LLMs to extract latent patterns from interaction chains, we convert them into structured interaction-of-thoughts using a \textbf{progressive}, \textbf{masked} prompting strategy, which encourages the model to generate step-by-step reasoning by building on prior outputs. Rather than distilling full reasoning traces, which risk invalid outputs and incur costly rejection sampling, we are inspired by the analogy to supervised learning, where optimizers map input-output pairs to parameter updates. Similarly, we treat the LLM as an implicit optimizer that, given a question-answer pair, learns to generate coherent reasoning traces aligned with user interaction behavior.

To realize this, we employ a masked prompting strategy, wherein the correct answer is injected into the prompt, but the LLM is instructed to ignore it explicitly and instead generate the intermediate reasoning steps as if the answer were unknown, thereby using the answer implicitly to guide the reasoning process without directly revealing it. Otherwise, the progressive prompting design instructs the LLM to generate reasoning in a step-by-step manner, where each subsequent reasoning step depends solely on the results of previous steps. Therefore, we can represent the generation of the reasoning process as:
\begin{equation}
    r=\mathcal{LLM}(\mathcal{S},\mathcal{Q},\mathcal{A})
\end{equation}
where $\mathcal{Q}$ refers to the predicted rating that the start user would assign to the final item in an unclosed forward interaction chain, $A$ represents the correct rating of the final item by the start user, and $S$ is the well-designed progressive, masked prompting strategy.

\subsection{LLM post-training pipeline}
\label{pipline}
\subsubsection{Supervised Fine-tuning}

\begin{figure}[ht]
\begin{center}
\includegraphics[width=1.0\linewidth]{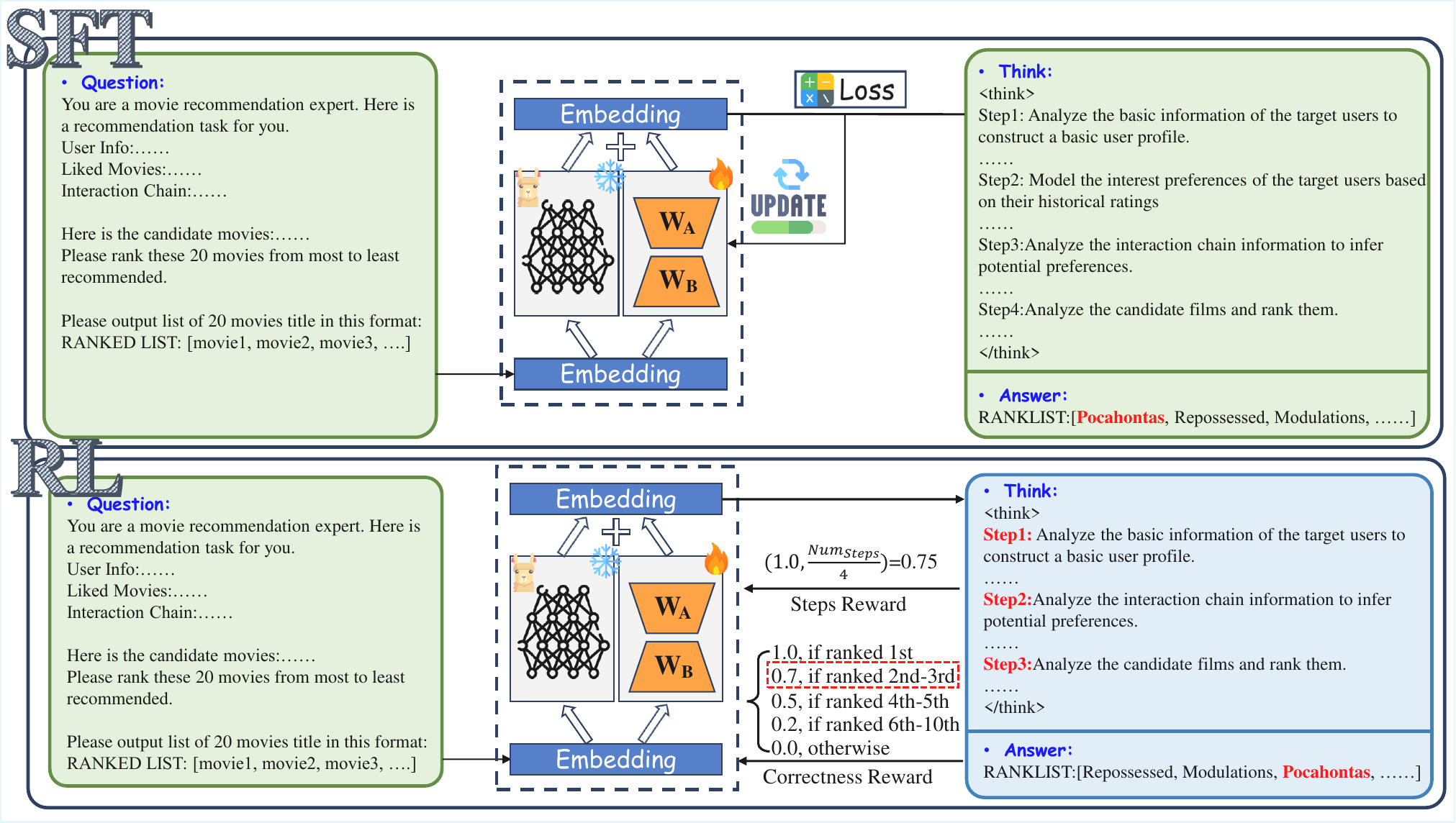}
\caption{The two-stage training pipeline for improving recommendation performance.}   
\label{Fig3}
\end{center}
\end{figure}

To equip the LLM with the ability to autonomously reason over interaction chains, we introduce a supervised fine-tuning~\cite{hu2021loralowrankadaptationlarge} stage that serves as a warm-up phase for reasoning-based recommendation, as depicted in Figure~\ref{Fig3}. Although our progressive, masked prompt design enables the generation of structured reasoning chains, producing such reasoning traces still requires external prompting or strong teacher LLMs, which limits scalability. To address this, we aim to internalize the reasoning process by training the LLM to generate reasoning chains on its own from a limited set of high-quality examples.
To construct training data, we adopt the masked prompt strategy described in Section ~\ref{reasoning chain}. Specifically, given a recommendation query $Q$, which asks the LLM to predict a user's preference for an item based on an interaction chain, and a correct answer $A$, we prompt the LLM to generate an intermediate reasoning chain $R$ that logically connects $Q$ and $A$. Importantly, while $A$ is embedded in the prompt, the LLM is instructed to ignore it explicitly and instead simulate the reasoning process as if it were solving the problem from scratch. This allows the generated reasoning to be guided yet faithful to the underlying interaction context. Each training instance is structured as a triplet $(Q, R, A)$, which is then used to optimize the LLM’s generation behavior through supervised learning. The training objective is formalized as:
\begin{equation}
    \theta^\ast = \arg\min_{\theta} \ \mathbb{E}_{(Q, R, A) \sim \mathcal{D}} \ \mathcal{L}(\text{LLM}_\theta(Q), R, A)
\end{equation}
Here, $\mathcal{D}$ denotes the constructed training dataset and $\mathcal{L}$ is the loss function encouraging the LLM to produce reasoning chains and answers aligned with the gold supervision. 
\subsubsection{Reinforcement Learning}
To further enhance the LLM’s ability to reason over interaction chains and improve its generalization beyond limited annotated examples, we introduce a reinforcement learning (RL) stage. This stage is designed to provide scalable supervision signals by allowing the LLM to refine its reasoning strategies through interaction with a reward function, without relying on additional labeled reasoning chains. In this way, RL complements the supervised fine-tuning phase by enabling continuous self-improvement through feedback. We adopt the Group Relative Policy Optimization (GRPO) algorithm~\cite{shao2024deepseekmathpushinglimitsmathematical}, which stabilizes policy updates by computing group-normalized advantages over multiple sampled completions without requiring a separate value network. The GRPO objective is defined as:
\begin{align}
\mathcal{J}_{\text{GRPO}}(\theta)
& =\mathbb{E}_{q \sim P(Q), \{o_i\}_{i=1}^{G} \sim \pi_{\theta_{\text{old}}}(O|q)} \\
&  \left[ \frac{1}{G} \sum_{i=1}^{G} \min \left( 
\frac{\pi_\theta(o_i|q)}{\pi_{\theta_{\text{old}}}(o_i|q)} A_i,\right. \right. \nonumber 
\left. \left. \text{clip} \left( 
\frac{\pi_\theta(o_i|q)}{\pi_{\theta_{\text{old}}}(o_i|q)}, 1 - \epsilon, 1 + \epsilon 
\right) A_i \right) 
- \beta D_{\text{KL}}(\pi_\theta \parallel \pi_{\text{ref}}) \right]
\end{align}
where $A_i$ denotes the group-relative advantage, computed by normalizing the rewards of actions within the same group, thereby stabilizing training and improving efficiency. To effectively guide the optimization, we design a reward function composed of two parts: a format reward called reasoning step reward to encourage appropriate reasoning length and avoid overly brief or verbose outputs, and a performance reward called ranking correctness reward that scores the recommendation quality based on the predicted rank of the ground-truth item. The final reward is computed as a weighted sum, assigning weight 1 to the reasoning step reward and weight 2 to the correctness reward. Full details and formulations are provided in Appendix~\ref{reward_design}.

\section{Experiments}
\label{experiment}
In this section, we evaluate R2Rec across multiple dimensions, including comparisons with classical and LLM-based baselines on three datasets, ablation studies on interaction chains and the reasoning-driven training pipeline (SFT and RL), cross-domain transfer experiments, and case studies demonstrating interpretable, reasoning-based recommendations.
\subsection{Experimental Settings}
\subsubsection{Datasets}
\label{dataset} 
\begin{itemize}[leftmargin=10pt]
    \item \textbf{MovieLens-1M}~\cite{10.1145/2827872} is a commonly-used movie recommendation dataset contains user demographics, movie attributes and user-movie ratings.
    \item \textbf{Amazon}~\cite{hou2024bridging} is derived from Amazon, a leading global e-commerce platform, which provides rich information including user reviews, item metadata, and user-item links. In our work, we select datasets from two domains: Book and Electronics.
\end{itemize}
For each dataset, we retain users with at least six ratings above 3 and sort their ratings chronologically. The most recent positively rated item is used as the ground-truth. We create a candidate set by combining the positive item with 19 randomly sampled negative items and evaluate the LLM’s ranking. Each dataset includes 1,000 randomly sampled users for evaluation.
\subsubsection{Implementation Details}
\label{LLM}
We use DeepSeek-R1-Distill-Llama-8B~\cite{guo2025deepseek} as the backbone LLM, with temperature set to 0.6 and top-p to 0.9. For each user, the prompt includes demographic features (if available), five recently liked items, five sampled interaction chains, and a 20-item candidate set. During the SFT stage, we apply Low-Rank Adaptation (LoRA)~\cite{hu2021loralowrankadaptationlarge} to fine-tune the LLM using the AdamW optimizer with a learning rate of 5e-5, for 10 epochs and 500 training samples. In the RL stage, we further train the LoRA adapter obtained from SFT with the same optimizer, using a smaller learning rate of 5e-6, for 1 epoch and 500 training samples. 

\subsection{Performance Comparsion}
In this section, we evaluate R2Rec against classical and LLM-based baselines on MovieLens, Amazon Book, and Amazon Electronics using HitRatio@1, HitRatio@3, and NDCG@3. (We omit NDCG@1 as it is equivalent to HitRatio@1 due to the single positive item in each candidate set.)
\subsubsection{Baselines}
\label{baseline}
\begin{itemize}[leftmargin=15pt]
    \item Classic Methods: \textbf{GraphSage}~\cite{hamilton2018inductiverepresentationlearninglarge}, \textbf{NGCF}~\cite{Wang_2019}, \textbf{LightGCN}~\cite{he2020lightgcnsimplifyingpoweringgraph} in graph neural networks for recommendation, as well as \textbf{GRU4Rec}~\cite{hidasi2016sessionbasedrecommendationsrecurrentneural}, \textbf{Caser}~\cite{tang2018personalizedtopnsequentialrecommendation} and \textbf{SASRec}~\cite{kang2018selfattentivesequentialrecommendation}, which are RNN-based, CNN-based and attention-based recommenders, respectively.
    \item LLM-based Methods: \textbf{GPT-4o-mini} and \textbf{o4-mini} are closed-source LLMs released by OpenAI. \textbf{ChatRec}~\cite{gao2023chatrecinteractiveexplainablellmsaugmented} reformulates user data into prompts for recommendations. \textbf{TALLRec}~\cite{Bao_2023} fine-tunes LLMs on instruction-style recommendation data. \textbf{LLaRA}~\cite{liao2024llaralargelanguagerecommendationassistant} incorporates user interaction sequences as an additional modality.
\end{itemize}
\label{RQ1}
\subsubsection{Results}
We present the experimental results in Table~\ref{Table1}, where each result is the average of three independent runs. The observation can be summarized as follows.
\begin{table}[ht]
\caption{The Results of R2Rec compared with classical recommendation models and LLMs-based methods. Bold and underline indicate the best and second-best performance.}
\label{Table1}
\centering
\resizebox{1.0\textwidth}{!}{
\Huge
\begin{tabular}{l@{}cccc|ccc|ccc}
\toprule
  &\multirow{2}{*}{Method}& \multicolumn{3}{c|}{MovieLens}& \multicolumn{3}{c|}{Amazon Book}& \multicolumn{3}{c}{Amazon Electronics}\\
                         && HitRatio@1&    HitRatio@3& NDCG@3& HitRatio@1&    HitRatio@3& NDCG@3& HitRatio@1&    HitRatio@3& NDCG@3\\\midrule
 \multirow{6}{*}{Classical} &GraphSage& 0.313& 0.491& 0.461& 0.321& 0.502& 0.465& 0.318& 0.516& 0.443\\
  &NGCF& 0.358& {\ul0.512}& {\ul0.472}& 0.455& 0.521& 0.497& 0.372& 0.524& 0.451\\
  &LightGCN& 0.276& 0.491& 0.435& 0.362& 0.537& 0.465& 0.379& {\ul0.563}& 0.487\\
  &GRU4Rec& 0.376& 0.504& 0.452& 0.495& 0.572& 0.527& 0.386& 0.521& {\ul0.498}\\
  &Caser& 0.319& 0.496& 0.421& 0.422& {\ul0.631}& 0.544& 0.385& 0.518& 0.496\\
  & SASRec& 0.380& 0.510& 0.452& {\ul0.531}& 0.592& 0.544& 0.391& 0.514&0.453\\\midrule
 \multirow{6}{*}{LLMs-based} &GPT-4o-mini& 0.240& 0.374& 0.317& 0.344& 0.502& 0.436& 0.222& 0.318& 0.276\\
  & o4-mini& 0.274& 0.419& 0.357& 0.466& 0.596& 0.541& 0.282& 0.372&0.335\\
  &ChatRec& 0.230& 0.340& 0.293& 0.304& 0.408& 0.362& 0.258& 0.370& 0.315\\
  &TallRec& 0.318& 0.442& 0.401& 0.412& 0.478& 0.421& 0.342& 0.432& 0.368\\
  &LLaRA& {\ul0.371}& 0.473& 0.454& 0.528& 0.583& {\ul0.556}& {\ul0.403}& 0.526& 0.491\\\midrule
 &R2Rec& \textbf{0.404}& \textbf{0.544} & \textbf{0.484} & \textbf{0.586} & \textbf{0.656} & \textbf{0.627} & \textbf{0.460}& \textbf{0.596}& \textbf{0.538}\\ \bottomrule
\end{tabular}}
\end{table}

As for classic methods, including GNN-based methods (e.g., GraphSAGE, NGCF, LightGCN) and deep neural models (e.g., GRU4Rec, Caser, SASRec), they all consistently underperform R2Rec. These models predominantly rely on observable behavioral patterns from user–item interactions, lacking the capacity to incorporate semantic item information or infer the underlying rationale behind user preferences. This performance gap underscores the significance of equipping recommendation systems with both semantic comprehension and reasoning abilities.

For LLM-based methods, we analyze them from two perspectives. First, closed-source LLMs achieve competitive results on certain datasets, highlighting the utility of their world knowledge and semantic understanding in recommendation. Notably, o4-mini, a reasoning-enhanced model, consistently outperforms GPT-4o-mini, suggesting that improved reasoning capabilities boost recommendation performance. Second, LLM4Rec approaches (e.g., ChatRec, TallRec, LLaRA) often surpass classical models by aligning LLMs with recommendation tasks and leveraging their semantic modeling strengths. However, their performance occasionally lags behind o4-mini, underscoring the importance of deeper reasoning over user-item interactions.

R2Rec outperforms all baseline models across the three datasets, achieving an average improvement of 10.48\% in HitRatio@1 compared to the second-best baseline. Specifically, R2Rec attains the highest HitRatio@1 scores of 0.404, 0.586, and 0.460 on MovieLens, Amazon-Book, and Amazon-Electronics, respectively, with relative improvements of 6.32\%, 10.99\%, and 14.14\% over the second-best baselines on each dataset. These results demonstrate that utilizing the reasoning ability of LLMs to mine the rich signals embedded in interaction chains can effectively enhance the recommendation performance of LLMs.
\subsection{Ablation Study}

\begin{table}[ht]
\caption{Ablation studies }
\label{Table2}
\centering
\resizebox{1.0\textwidth}{!}{
\Huge
\begin{tabular}{l@{}cccc|ccc|ccc}
\toprule
 &\multirow{2}{*}{Method}& \multicolumn{3}{c|}{MovieLens}& \multicolumn{3}{c|}{Amazon Book}& \multicolumn{3}{c}{Amazon Electronics}\\
                        && HitRatio@1&    HitRatio@3& NDCG@3& HitRatio@1&    HitRatio@3& NDCG@3& HitRatio@1&    HitRatio@3& NDCG@3\\ \midrule
 \multirow{3}{*}{Prompt}&Full& 0.230& 0.381& 0.317& 0.366& 0.494& 0.440& 0.192& 0.326& 0.268\\
 &w/o Reasoning Chain& 0.186& 0.327& 0.267& 0.278& 0.420& 0.361& 0.172& 0.256& 0.219\\
 &w/o Interaction Chain& 0.156& 0.299& 0.237& 0.160& 0.324& 0.255& 0.072& 0.156& 0.118\\\midrule
 \multirow{3}{*}{Training Stage}& Full& 0.404& 0.544 & 0.484 & 0.586 & 0.656 & 0.627 & 0.460& 0.596&0.538\\
  &w/o RL& 0.320& 0.453& 0.397& 0.528& 0.608& 0.574& 0.436& 0.490&0.467\\
  &w/o SFT& 0.156& 0.270& 0.221& 0.248& 0.340& 0.300& 0.112& 0.158&0.139\\\bottomrule
\end{tabular}}
\end{table}
In order to verify the validity of our designs, we conduct ablation studies regarding both interaction chain features in prompt and the stages of training pipeline. We show the results in Table~\ref{Table2}.
\subsubsection{Ablation Study of Interaction Chain}
The upper block of Table~\ref{Table2} investigates the impact of removing components from the prompt. We observe that eliminating the reasoning chain leads to consistent performance drops across all datasets, indicating that explicitly modeling reasoning is beneficial. More notably, removing the interaction chain results in the most substantial degradation—e.g., on Amazon Electronics, HitRatio@1 drops from 0.192 to 0.072—suggesting that the interaction chain serves as the essential substrate for reasoning. In other words, the interaction chain provides the behavioral context from which reasoning chains can be derived. Without this foundation, the LLM lacks the necessary signals to support preference inference.

\subsubsection{Ablation Study of Training Stage}
The bottom part of Table~\ref{Table2} shows the impact of removing training stages: omitting supervised fine-tuning (SFT) causes a significant drop in performance (e.g., HitRatio@1 on MovieLens falls from 0.404 to 0.156), highlighting SFT’s critical role in warm-starting the model’s reasoning ability using limited curated data. Removing reinforcement learning (RL) also degrades results, though less severely, demonstrating RL’s value in providing scalable supervision by enabling the LLM to refine reasoning via reward feedback without labeled data. Further, Figure~\ref{reward} compares reward curves during RL training with and without SFT, showing that the SFT-initialized model achieves higher and more stable rewards, while training from scratch leads to instability and suboptimal convergence (detailed analysis is in Appendix~\ref{reward_curve}).

\subsection{Transferability}

\begin{table}[ht]
\caption{Transferability Study}
\label{Table3}
\centering
\resizebox{1.0\textwidth}{!}{
\Huge
\begin{tabular}{l@{}cccc|ccc|ccc}
\toprule
  && \multicolumn{9}{c}{Domain of Test Data}\\
\midrule
 && \multicolumn{3}{c|}{MovieLens}& \multicolumn{3}{c|}{Amazon Book}& \multicolumn{3}{c}{Amazon Electronics}\\
 & & HitRatio@1& HitRatio@3& NDCG@3& HitRatio@1& HitRatio@3& NDCG@3& HitRatio@1& HitRatio@3& NDCG@3\\\midrule
                        \multirow{3}{*}{Domain of Training Data}&MovieLens& 0.404&    0.544& 0.484& 0.502&    0.616& 0.540& 0.302&    0.362& 0.336\\ 
 &Amazon Book& 0.302& 0.442& 0.381& 0.586& 0.656& 0.627& 0.430& 0.478& 0.458\\
 &Amazon Electronics& 0.238& 0.324& 0.284& 0.484& 0.594& 0.547& 0.460& 0.596& 0.538\\ \bottomrule
\end{tabular}}
\end{table}
To evaluate the transferability of our method, we perform cross-domain experiments by training on one dataset and testing on another. The results are shown in Table~\ref{Table3}. For example, the LLM trained on MovieLens achieves an HitRatio@1 of 0.502 when tested on Amazon Book, which is competitive and even outperforms several in-domain baselines. This indicates that the LLM acquires a transferable reasoning ability over interaction chains, rather than overfitting to domain-specific user preferences. Such capability allows it to generalize across different domains by leveraging structural patterns in user-item interactions rather than relying solely on memorized user behavior.

\subsection{Case Study}
\vspace{0mm}
\begin{figure}[ht]
\begin{center}
\includegraphics[width=1.0\linewidth]{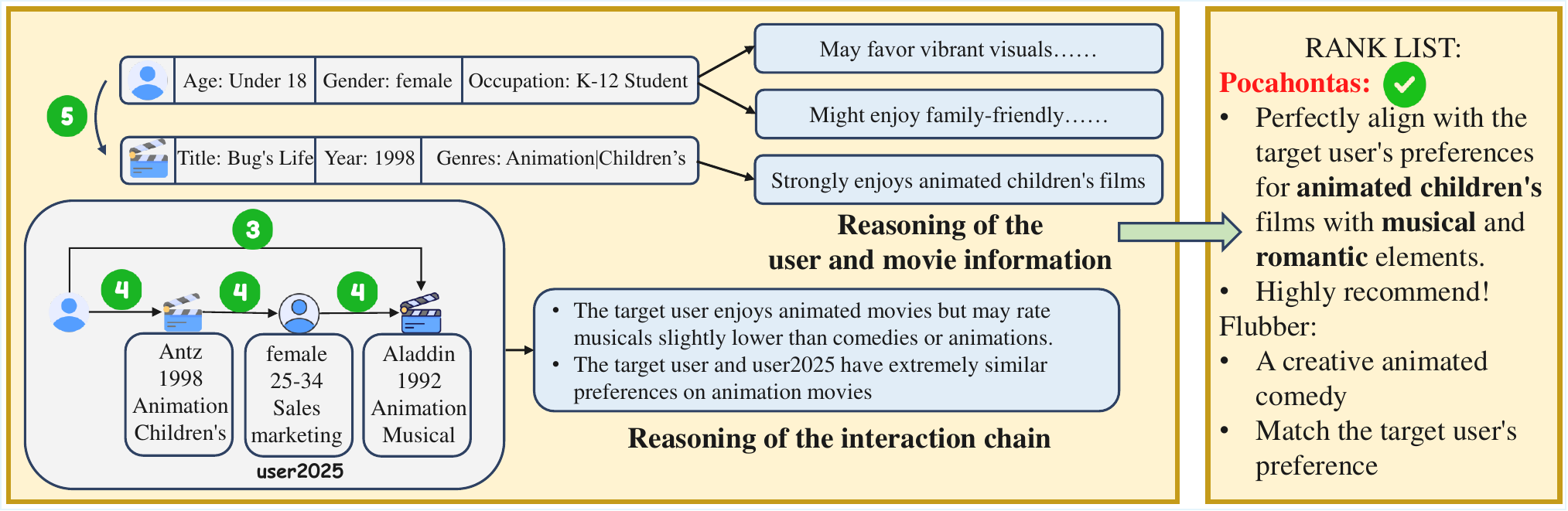}
\caption{Case Study}   
\label{Fig4}
\end{center}
\end{figure}

To illustrate the interpretability of our approach, we present a case study involving a female user under 18 years old, as shown in Figure~\ref{Fig4}. Based on her demographic profile and a high rating for A Bug’s Life (1998), the model infers a preference for animated, children-oriented comedies. Reasoning over an interaction chain involving shared ratings of Antz and Aladdin with a similar user further refines this understanding, suggesting a nuanced preference for animation and lighthearted themes over musicals. Given this profile, the model ranks Pocahontas (1995)—an animated musical with children’s and romantic elements—as the most suitable recommendation. This example highlights how our method enables interpretable recommendations by explicitly tracing the reasoning from both personal history and interaction structure.
\section{Conclusions}
\label{conclusion}
In this paper, we propose a reasoning-enhanced recommendation method that introduces interaction-of-thoughts—structured reasoning sequences derived from user–item interaction chains—to guide LLMs in modeling user preferences. To internalize this reasoning process, we design a two-stage training pipeline that combines supervised fine-tuning with reinforcement learning. Extensive experiments across diverse recommendation domains demonstrate significant performance gains, improved generalization, and enhanced interpretability. A key limitation of our method is the constrained context length of LLMs, which restricts the number of interaction-of-thought sequences that can be processed simultaneously, potentially limiting the model’s effectiveness. Future work may explore retrieval-augmented prompting and more compact encoding methods to overcome these challenges and further scale reasoning-aware recommendation.

\bibliographystyle{unsrt}
\bibliography{reference}
\newpage
\appendix
\section{Appendix}
\subsection{The progressive, masked prompting strategy to generate Interaction-of-Thoughts}
The specific prompt to generate Interaction-of-thoughts is as follows(Using the Movielens dataset as an example).

\textbf{Prompt}: You are a movie recommendation expert. I will give you a user-movie interaction chain, a question and a correct answer. The interaction chain contains the user's rating of a certain movie and the basic information of the user/movie. Movie ratings are integers from 1 to 5, with 5 standing for very liked, 4 standing for liked, 3 standing for neutral, 2 standing for disliked and 1 standing for very disliked. The following is the form of the interaction chain that shows the common viewing behavior and ratings among users, and the basic information of the user/movie is also recorded in this interaction chain:
(Target user)(Basic information of Target user) -- (Rating 5) -- (Movie A)(Basic information of Movie A) -- (Rating 4) -- (User 1234)(Basic information of User1234) -- (Rating 2) -- (Movie B)(Basic information of Movie B) -- (Rating 3) -- (Target user)
The meaning of the above interaction chain is:
The target user has watched the Movie A and gave it a score of 5, while user1234 has also watched Movie A and gave it a score of 4, and user1234 has also watched the Movie B and gave it a score of 2. This target user has watched the Movie B as well and gives it a score of 3.
Your task is to pretend to reason step by step about this interaction chain without knowing the given answer in advance, independently analyze the information of each jump in the interaction chain, and gradually obtain the final answer. The reasoning process must comply with the following requirements:
1. Your reasoning process should be progressive, and the next step of reasoning should only rely on the previous reasoning results.
2. Your reasoning process should be masked. When reasoning about the information of a certain hop in the interaction chain, the only information you can utilize is the information before that hop, and you cannot utilize the information after that hop.
3. Your reasoning process should not contain any information about the final answer. The answers provided are only to guide your reasoning process.
4. The answer obtained based on your reasoning process should be consistent with the correct one.

Next, I will provide you with a user-movie interaction chain, a question and a correct answer:

Interaction chain:
\textit{\{Interaction Chain\}}

Question: 
What is the rating of the movie \textit{\{Final movie of the interaction chain\}} given by the target users?

Answer: 
Rating \textit{\{True rating of the target user given to the final movie\}}

Please output your reasoning process and the final answer obtained based on the reasoning process:

Reasoning process:

Final answer:
\subsection{Reward Function Design}
\label{reward_design}

We define the total reward used in RL training as a weighted combination of two components: the reasoning step reward and the ranking correctness reward.

\textbf{Reasoning Step Reward.}  
This reward encourages the model to generate an appropriate number of reasoning steps—neither too short nor unnecessarily long—based on the number of reasoning statements identified in the response:
\begin{equation}
\text{reward}_{\text{step}} = \min\left(1.0, \frac{Num_{\text{steps}}}{4} \right)
\end{equation}

\textbf{Ranking Correctness Reward.}  
This reward reflects how well the ground-truth item is ranked in the output list. A higher score is assigned when the ground-truth item is ranked higher:
\begin{equation}
\text{reward}_{\text{correctness}}=
\begin{cases}
1.0,& \text{if ranked 1st}\\
0.7,& \text{if ranked 2nd-3rd}\\
0.5,& \text{if ranked 4th-5th}\\
0.2,& \text{if ranked 6th-10th}\\
0.0,& \text{otherwise}
\end{cases}  
\end{equation}

\textbf{Final Reward.}  
The final reward used for optimization is calculated as:
\begin{equation}
\text{reward}_{\text{total}} = 1 \cdot \text{reward}_{\text{step}} + 2 \cdot \text{reward}_{\text{correctness}}
\end{equation}

\subsection{Reward Curve Analysis During RL Training}
\label{reward_curve}

To better understand the contribution of supervised fine-tuning (SFT) in initializing the LLM’s reasoning capabilities, we analyze the reward trajectories during reinforcement learning (RL) training for models with and without SFT initialization.

\begin{figure}[ht]
\begin{center}
\includegraphics[width=1.0\linewidth]{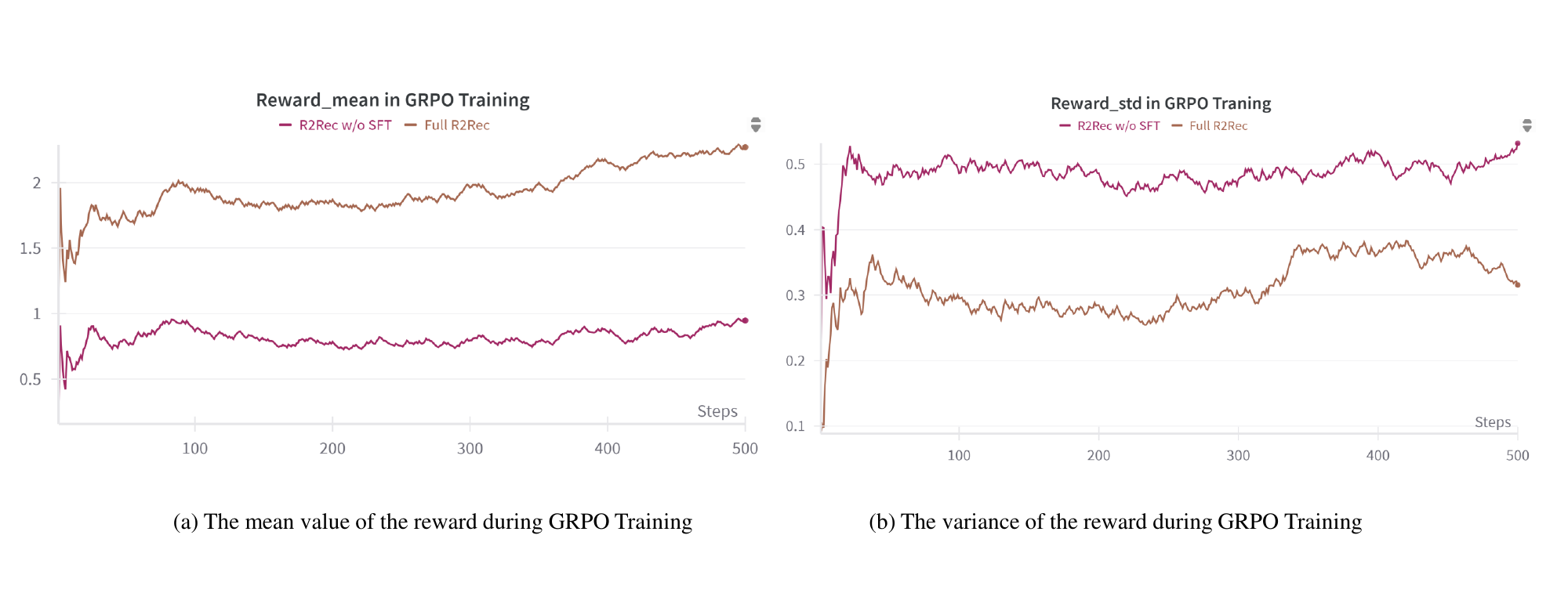}
\caption{Reward comparison between RL training with and without SFT initialization.}
\label{reward}
\end{center}
\end{figure}

As shown in Figure~\ref{reward}, the LLM initialized with SFT achieves a significantly higher starting reward and maintains stable reward growth throughout the training process. This indicates that SFT equips the model with a fundamental ability to reason over user–item interaction chains, allowing it to benefit more effectively from reward signals during RL.

In contrast, the model trained directly with RL from scratch exhibits lower and more volatile rewards. It converges more slowly and fails to reach the performance of the SFT-initialized counterpart. These findings highlight that supervised fine-tuning is a critical component for stabilizing RL training, especially in domains lacking abundant reasoning supervision, as it provides the model with essential prior structure to build upon during reward-driven optimization.

\subsection{Compute Resources}
\label{appendix:compute}

All experiments were conducted using NVIDIA A100-80G GPUs. For supervised fine-tuning (LoRA-based SFT), we used a single A100 GPU, with each training run taking approximately 2.5 hours. For reinforcement learning with the GRPO algorithm, we utilized two A100 GPUs, and each training run took approximately 8 hours. Evaluation and inference tasks were conducted using a single A100 GPU, with average inference time per user about 2 seconds.

\subsection{Ethical Considerations}
\label{appendix:ethics}

All datasets used in this study—MovieLens, Amazon Books, and Amazon Electronics—are publicly available and anonymized. They contain no personal or identifiable information and are widely used in the recommendation systems research community. 

No human subjects were involved in this study, and no private user data was collected or processed. This work does not pose foreseeable risks related to privacy, fairness, or safety, and complies fully with the NeurIPS Code of Ethics.

\end{document}